\documentclass[journal,twoside]{asmb}

\begin{document}
\title{RIS-Aided Free-Space Optics Communications in A2G Networks over Inverted Gamma-Gamma  Turbulent Channels}

\author{
\IEEEauthorblockN{Md. Abdur Rakib, Md. Ibrahim,
\textit{Graduate Student Member, IEEE}, A. S. M. Badrudduza, \textit{ Member, IEEE}, Imran Shafique Ansari, \textit{Senior Member, IEEE},  Md. Shahid Uz Zaman and Heejung Yu,
\textit{Senior Member, IEEE}}}

\twocolumn[
\begin{@twocolumnfalse}
\maketitle
\begin{abstract}
\section*{Abstract}

With the advent of sixth-generation networks, reconfigurable intelligent surfaces (RISs) have revolutionized wireless communications through dynamic electromagnetic wave manipulation, thereby facilitating the adaptability and unparalleled control of real-time performance evaluations. This study proposed a framework to analyze the performance of RIS-assisted free-space optics (FSO) communication over doubly inverted Gamma-Gamma (IGGG) distributions with pointing error impairments. Furthermore, a special scenario addressing secure communication in the potential presence of an eavesdropper. Consequently, we derived closed-form expressions for the outage probability, average bit error rate, average channel capacity, average secrecy capacity, and secrecy outage probability by employing an asymptotic analysis to provide deeper insights into the influence of various system parameters. Finally, we verified our analytical results through appropriate numerical simulations.
\end{abstract}

\begin{IEEEkeywords}
\section*{Keywords} 

Reconfigurable intelligent surface, free-space optics, air-to-ground network, inverted Gamma-Gamma channels, secrecy outage probability
\end{IEEEkeywords}
\end{@twocolumnfalse}
]
\section{INTRODUCTION}
Owing to the advent of free-space optics (FSO) wireless communications, a revolutionary era for optical communications has been ushered in through the facilitation of impressive data transmission speeds and versatile applications, including backhaul support and disaster recovery \cite{art9}. However, FSO is plagued by several challenges over extended distances, primarily because of its sensitivity to atmospheric conditions and pointing errors. To address this issue, reconfigurable intelligent surfaces (RISs), that utilize passive reflecting components to construct smart wireless environments, are among the most effective solutions \cite{ris1, ris2, ris3, ris4}. The optimization of the signal phase and amplitude via the RIS alleviates adverse effects and enhances the secrecy performance in FSO wireless communication, particularly in the context of potential eavesdropping risks \cite{ris-pls1, ris-pls2, ris-pls3}. The rapid progress in sixth-generation ($6$G) wireless networks has garnered the attention of the research community \cite{6g}. Recently, several studies \cite{art12,art14,art13} have evaluated the performance of RIS-assisted FSO models. In \cite{art12}, the authors investigated the performance of RIS-enabled FSO systems, incorporating a combination of M\'alaga $(\mathcal{M})$, Fisher–Snedecor $(\mathcal{F})$, and Gamma-Gamma (GG) distributions. Recently, an RIS-empowered $\mathcal{F}$ turbulent model was proposed in \cite{art14}, wherein the closed-form expressions for the ergodic capacity (EC), outage probability (OP), and average bit error rate (ABER) were derived to evaluate the system performance. Another study in \cite{art13} concluded that RIS-aided wireless systems with a substantial number of reflecting elements exhibited better performance than relay-aided systems. Notably, the investigation of the secrecy performance within RIS-assisted FSO systems has not received substantial attention in previous studies \cite{art12,art14,art13}. The existing literature predominantly has focused on the assessment of system performance under $\mathcal{M}$ and GG turbulent distributions, whereas the investigation of doubly inverted Gamma-Gamma (IGGG) distributions remains an open problem.

In \cite{f1,f2,f3}, extensive analyses of physical layer security (PLS) in FSO communications were conducted. The authors of \cite{f1} analyzed the secrecy performance by assuming various wiretapping situations based on the location of the eavesdropper. In \cite{f3},
the effects of atmospheric turbulence and pointing error on the secrecy performance were investigated over $\mathcal{M}$ distributions. However, the study in \cite{f1,f2,f3} did not examine the influence of RIS in the case of secured FSO transmission. Studies on PLS performance over RIS-aided FSO systems are scarce \cite{pls_ed}. {Recently, \cite{art16} inspected the impact of RIS reflecting elements over Rayleigh fading channels, where the authors derived analytical expressions of the secrecy outage probability (SOP) to evaluate secrecy performance.}

The authors of \cite{art17} recently introduced the IGGG model, which acts as a generalized representation of both $\mathcal{F}$ and GG distributions. Although this model demonstrates superior compatibility with experimental data compared to the log-normal, inverse Gaussian Gamma (IGG), $\mathcal{F}$, and GG models, only one study by \cite{art1} has assessed its performance. {The introduction of RIS into FSO channels extends the coverage area while enhancing the secrecy performance. According to the current literature, most researchers consider $\mathcal{M}$ \cite{10313311} and $\mathcal{F}$ \cite{stefanovic2021performance} channels for FSO communications. To the best of our knowledge, no prior research has addressed the incorporation of RIS into an IGGG distributed turbulent model.} Moreover, the existing literature lacks investigations on the secrecy performance of RIS-assisted FSO system modeling with the IGGG channel, particularly in scenarios wherein there exists a risk of interruption of confidential data. Thus, focusing on these unexplored realms of investigation, this study proposed a wireless network of RIS-assisted FSO communication over a doubly IGGG turbulent system. The key contributions of this study are outlined as follows:

\begin{itemize}
    \item  We proposed an RIS-based IGGG turbulent model for FSO communications for air-to-ground (A2G) networks. In contrast to previous studies that have solely focused on performance analysis, we proposed a comprehensive PLS framework that considered the presence of an eavesdropper. {Unlike \cite{art17}, our primary focus was the investigation of the impact of RIS on the system performance. Furthermore, we evaluated the proposed model in the presence of an eavesdropper. Thus, the mathematical analysis conducted in this study is distinctly different in terms of system configuration.}
    
    \item The probability density function (PDF) and cumulative density function (CDF) of the signal-to-noise ratio (SNR) in the proposed system model were derived. Therefore, utilizing these expressions, the analytical expressions of OP, ABER, average channel capacity (ACC), average secrecy capacity (ASC), and SOP were obtained in closed-form expressions. Furthermore, certain numerical outcomes were observed based on these mathematical expressions. In addition, asymptotic analyses were conducted under high SNR conditions for additional insights.

    \item To assess the system performance, this study examined the influence of diverse factors such as atmospheric turbulence conditions, receiver detection techniques, pointing error impairments, and the number of reflecting elements in RIS. The accuracy of the analytical results was validated through numerical simulations.
\end{itemize}


The remainder of this paper is structured as follows: Section \ref{sm} introduces the proposed model. Section \ref{sa} presents the statistical analysis results of PDF and CDF. The analytical expressions for various performance metrics (OP, ABER, ACC, ASC and SOP) are derived in Section \ref{pa}. Section \ref{results} presents the numerical results and key observations. Finally, Section \ref{co} concludes the study.

\section{SYSTEM MODEL AND PROBLEM FORMULATION }
\label{sm}
\begin{figure*}[t]
   \vspace{-5mm}
       \centerline{\includegraphics[width=0.6\textwidth,angle =0]{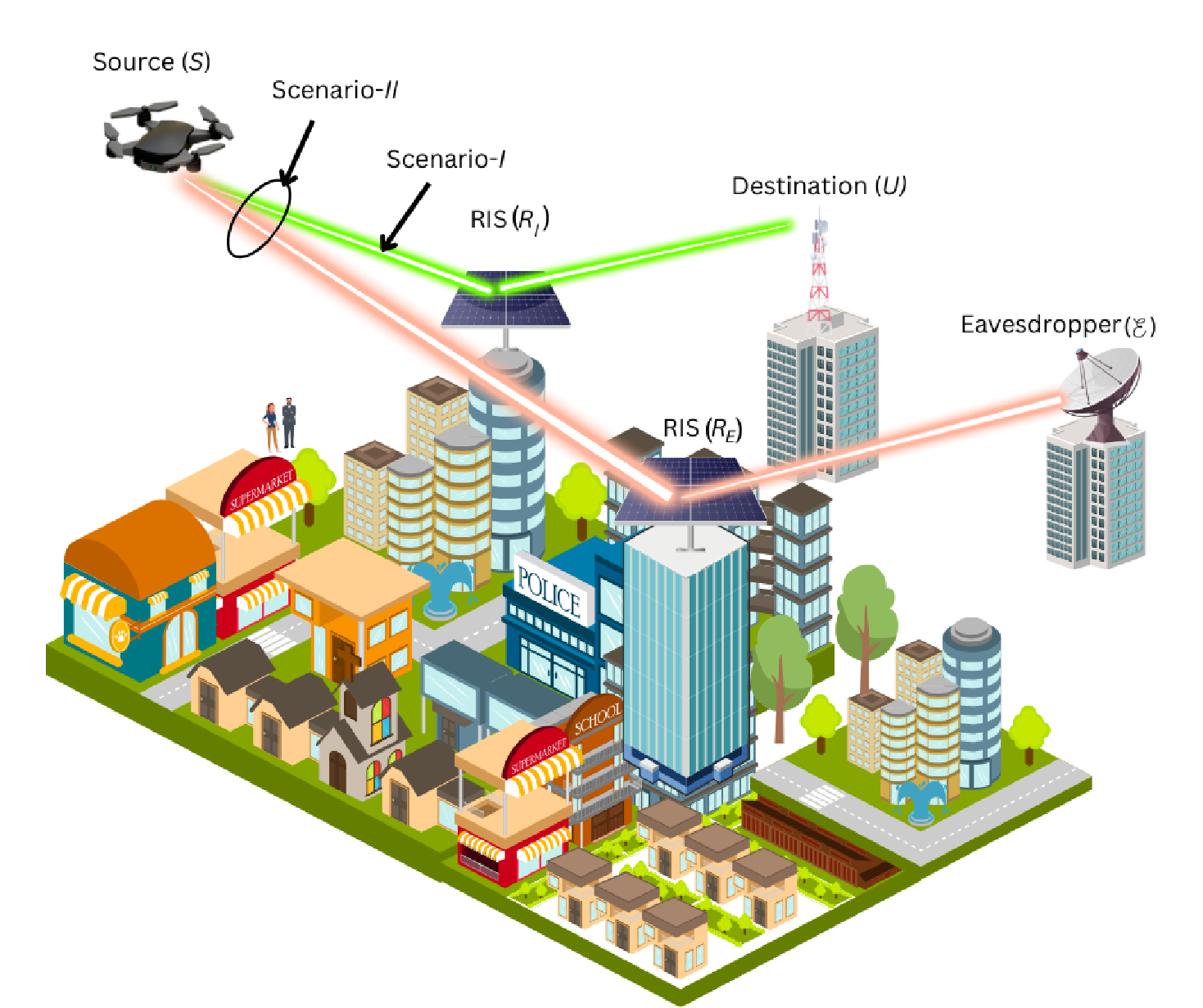}}
     \caption{{System configuration comprising a source ($\mathcal{S}$), two RISs ($\mathcal{R_{I}}$ and $\mathcal{R_{E}}$), a destination user ($\mathcal{U}$), and an eavesdropper ($\mathcal{E}$).}}
       \label{fig1}
   \end{figure*}
The proposed system model representing a doubly IGGG turbulent channel for FSO communication demonstrates the interaction between the source, $\mathcal{S}$ (airborne), and destination user, $\mathcal{U}$ (ground station). The connection overcomes obstacles between them by utilizing an RIS, $\mathcal{R_{I}}$ (implemented on top of the building) with $N_{d}$ reflecting elements (Scenario-\textit{I}). It is also assumed that an unauthorized party, referred to as an eavesdropper, $\mathcal{E}$ (smartphone, airborne), aims to intercept confidential information transmitted from $\mathcal{S}$ to $\mathcal{U}$ through a separate RIS, $\mathcal{R_{E}}$ with $N_e$ reflecting elements (Scenario-\textit{II}). Therefore, the received signal expression at $\mathcal{U}$ and $\mathcal{E}$ can be written as follows:
\begin{align}
    \label{s1}
y_{j}=\left[\sum_{t=1}^{N_{j}}f_{t,j}e^{i\Theta_{t,j}}g_{t,j}\right]y+w_{j},
\end{align}
where $j \in \{d,e\}$,
$d$ and $e$ correspond to $\mathcal{S}-\mathcal{R}_{\mathcal{I}}-\mathcal{U}$ and $\mathcal{S}-\mathcal{R}_{\mathcal{E}}-\mathcal{E}$ links, respectively, the channel gains for the first and second hops are denoted by $f_{t,j}=\rho_{t,j}e^{i\phi_{t,j}}$ and  $g_{t,j}=\varrho_{t,j}e^{i\varphi_{t,j}}$. Here, $\rho_{t,j}$ and $\varrho_{t,j}$ represent the IGGG distributed random variables (RVs), where $\phi_{t,j}$ and $\varphi_{t,j}$ denote the phases of the channel gains. The signal transmitted from $\mathcal{S}$ with signal power $P_{y}$ is denoted as $y$. The additive white Gaussian noise (AWGN), with noise power $P_{j}$, is represented by $w_{j}\sim\Tilde{\mathcal{N}}(0,P_{j})$. The matrix representation of \eqref{s1} is expressed as $y_{j}=\textbf{g}_{j}^{T}
\Theta_{j} \textbf{f}_{j}y+w_{j}$, where $\textbf{f}_{j}=[f_{1,j}  f_{2,j} \ldots f_{N_{j},j}]^{T}$ and $\textbf{g}_{j}=[g_{1,j} g_{2,j} \ldots g_{N_{j},j}]^{T}$ depict the channel coefficient vectors. Further, $\Theta_{j}$=diag$([e^{i\Theta_{1,j}} e^{i\Theta_{2,j}} \ldots e^{i\Theta_{N_{j},j}}])$ is the diagonal matrix incorporating phase transition utilized by RIS components. The expression for instantaneous SNR at $\mathcal{U}$ and $\mathcal{E}$ can be expressed as $\gamma_{j}=\left[\sum_{t=1}^{N_{j}}\rho_{t,j}\varrho_{t,j}e^{i(\Theta_{t,j}-\phi_{t,j}-\varphi_{t,j})}\right]^{2}\Bar{\gamma_{j}}$, where the average SNRs for $\mathcal{S}-\mathcal{R_{I}}-\mathcal{U}$ and $\mathcal{S}-\mathcal{R_{E}}-\mathcal{E}$ links can be denoted as $\Bar{\gamma_{j}}=\frac{P_{y}}{P_{j}}$. The phases $\mathcal{R_{I}}$ and $\mathcal{R_{E}}$ are controlled by the destination and eavesdropper to maximize the received signal power, respectively. Therefore, the maximization of instantaneous SNR can be obtained by setting  $\Theta_{t,j}=\phi_{t,j}+\varphi_{t,j}$ as $\gamma_{j}=\left(\sum_{t=1}^{N_{j}}\rho_{t,j}\varrho_{t,j}\right)^{2}\Bar{\gamma_{j}}$. 
\section{STATISTICAL ANALYSIS}
\label{sa}
\label{system}
\begin{lemma}
The PDF and CDF of $\gamma_{j}$ are expressed  as
\begin{align}
\label{e1}
f_{\gamma_{j}(\gamma)}&=\frac{\Lambda_{1,j}}{r\mu_{r,j}^{\frac{l_{j}}{r}}}\gamma^{\frac{l_{j}}{r}-1}G_{0,1}^{1,0}\left[\Lambda_{2,j}\left(\frac{\gamma}{\mu_{r,j}}\right)^{\frac{1}{r}}\biggl | 
    \begin{array}{c}
     - \\
     0 \\ 
    \end{array}
    \right],
\\
\label{e2}
F_{\gamma_{j}}(\gamma)&=\frac{\Lambda_{3,j}}{\sqrt{r}\mu_{r,j}^{\frac{l_{j}}{r}}}\gamma^{\frac{l_{j}}{r}}G_{1,r+1}^{r,1}\left[\Lambda_{4,j}\left(\frac{\gamma}{\mu_{r,j}}\right)\biggl | 
    \begin{array}{c}
     1-\frac{l_{j}}{r} \\
     0,{\frac{r-1}{r}},\frac{-l_{j}}{r} \\ 
    \end{array}
    \right],
\end{align}
where $\Lambda_{1,j}=\frac{\mathbb{E}(M_{t,j})^{l_{j}}}{\Gamma(l_{j})k_{j}^{l_{j}}}$, $\Lambda_{2,j}=\frac{\mathbb{E}(M_{t,j})}{k_{j}}$, $\Lambda_{3,j}=\frac{\mathbb{E}(M_{t,j})^{l_{j}}}{\Gamma(l_{j})k_{j}^{l_{j}}(2\pi)^{\frac{r-1}{2}}}$, $\Lambda_{4,j}=\left[\frac{\mathbb{E}(M_{t,j})}{k_{j}r}\right]^{r}$, $l_{j}=\frac{N_{j}\mathbb{E}(M_{t,j})^{2}}{\mathrm{Var}(M_{t,j})}$,
$k_{j}=\frac{\mathrm{Var}(M_{t,j})}{{\mathbb{E}(M_{t,j})}}$, 
electrical SNR is denoted as $\mu_{r,j}=\frac{[\eta_{j}\mathbb{E}(M_{t,j})]^{r}}{T_{0}}$, $T_{0}$ denotes the noise power, $r$ denotes the detection technique (i.e., $r=1$ defines the heterodyne detection (HD) detection and $r=2$ represents the intensity modulation/direct detection (IM/DD) detection), $\eta_{j}$ denotes the detector efficiency, and $G_{.,.}^{.,.}[\cdot]$ defines the Meijer's G function \cite[Eq.~(8.2.1.1)]{art3}.
\end{lemma}
\textit{Proof}: See Appendix.
\section{PERFORMANCE ANALYSIS}
\label{pa}
In this section, the analytical expressions of OP, ABER, ACC, ASC, and SOP are presented in closed-form. Moreover, an asymptotic analysis is performed to observe the effect of a high SNR owing to the proposed model.
\subsection{Outage Probability}
OP defines the probability that the system communication is disrupted when the output SNR falls below a predetermined threshold ($\gamma^{*}$). Therefore, the OP can be expressed as \cite[Eq.~(39)]{ar}
\begin{align}
    \label{op}   \mathrm{OP}=\mathrm{P_{r}}\left[\gamma_{d}\leq\gamma^{*}\right] =F_{\gamma_{d}}(\gamma^{*}).
\end{align}
\textit{Asymptotic Analysis:} By employing \cite[Eq.~(8.2.2.14)]{art3} to reverse the argument of Meijer's G function and subsequently applying \cite[Eq.~(41)]{art8} to expand it, the asymptotic expression of OP at a high SNR is obtained as
\begin{align} 
\label{ops}
    \mathrm{OP}^{\infty} \approx & \sum_{h=1}^{r}\frac{\Lambda_{3,d}}{\sqrt{r}\Lambda_{4,d}^{t_{1,h}-1}}\left(\frac{\gamma}{\mu_{r,d}}\right)^{\frac{l_{d}}{r}-t_{1,h}+1}
    \nonumber
    \\
\times & \frac{\prod_{g=1;g\neq h}^{r}\Gamma(t_{1,h}-t_{1,g})\Gamma(1+\frac{l_{d}}{r}-t_{1,h})}{\Gamma(1+t_{1,r+1}-t_{1,h})}, 
\end{align}
where $t_{1}=(1,\frac{1}{r},1+\frac{l_{d}}{r})$. Here, $t_{m,n}$ represents the $n^{th}$-term of $t_{m}$.

\subsection{Average Bit Error Rate }
ABER is a performance metric that quantifies the percentage of bits with errors compared to the total number of bits received during transmission. The mathematical representation of ABER as per various modulation techniques is defined as \cite[Eq.~(24)]{art6}
\begin{align}
    \label{ber_def}
    \mathrm{ABER}=\frac{q^{p}}{2\Gamma(p)}\int_{0}^{\infty}e^{-q\gamma}\gamma^{p-1}F_{\gamma_{d}}(\gamma)d\gamma,
\end{align}
where ($p$, $q$) denotes various modulation schemes. Substituting \eqref{e2} into \eqref{ber_def}, we obtain
\begin{align}
    \mathrm{ABER}=&\frac{q^{p}\Lambda_{3,d}}{\sqrt{r}\mu_{r,d}^{\frac{l_{d}}{r}}2\Gamma(p)}\int_{0}^{\infty}e^{-q\gamma}\gamma^{\frac{l_{d}}{r}+p-1}
    \nonumber
    \\
\times& G_{1,r+1}^{r,1}\left[\Lambda_{4,d}\left(\frac{\gamma}{\mu_{r,d}}\right)\biggl | 
    \begin{array}{c}
     1-\frac{l_{d}}{r} \\
     0,{\frac{r-1}{r}},\frac{-l_{d}}{r} \\ 
    \end{array}
    \right]d\gamma.
\end{align}
Employing the integral identity of \cite[Eq.~(8.4.3.1)]{art3} to transform $e^{-q\gamma}$ into Meijer's G function and subsequently applying \cite[Eq.~(2.24.1.1)]{art3}, the closed form expression of ABER is evaluated as
\begin{align}
\label{ber}
    \mathrm{ABER}=&\frac{\Lambda_{3,d}q^{\frac{-l_{d}}{r}}}{\sqrt{r}2\Gamma(p)\mu_{r,d}^{\frac{l_{d}}{r}}}G_{2,r+1}^{r,2}\left[\frac{\Lambda_{4,d}}{q\mu_{r,d}}\biggl | 
    \begin{array}{c}
     1-p-\frac{l_{d}}{r}, 1-\frac{l_{d}}{r} \\
     0,\frac{r-1}{r},\frac{-l_{d}}{r} \\ 
    \end{array}
    \right].
\end{align}
\textit{Asymptotic Analysis:} Utilizing the similar identity as employed in (\ref{ops}), the ABER is expressed asymptotically as
\begin{align}
\label{bers}
\mathrm{ABER}^{\infty} \approx & \sum_{h=1}^{r}\frac{\Lambda_{3,d}(q\mu_{r,d})^{t_{1,h}-\frac{l_{d}}{r}-1}}{\sqrt{r}2\Gamma(p)\Lambda_{4,d}^{t_{1,h}-1}}
\nonumber
\\
\times&\frac{\prod_{g=1;g\neq h}^{r}\Gamma(t_{1,h}-t_{1,g})\prod_{g=1}^{2}\Gamma(1+t_{2,g}-t_{1,h})}{\Gamma(1+t_{1,r+1}-t_{1,h})},
\end{align}
where $t_{2}=(p+\frac{l_{d}}{r},\frac{l_{d}}{r}).$

\subsection{Average Channel Capacity}
The ACC for a system, applicable to both HD and IM/DD techniques, is defined by \cite[Eq.~(29)]{art6}
\begin{align}
    \label{erg}
    \mathrm{ACC}=\frac{1}{2\ln{(2)}}\int_0^\infty\ln({1+\gamma})f_{\gamma_{d}}(\gamma)d\gamma.
\end{align}
Substituting \eqref{e1} into \eqref{erg}, we obtain
\begin{align}
    \label{erg2}
    \mathrm{ACC}&=\frac{\Lambda_{1,d}}{2\ln{(2)}r\mu_{r,d}^{\frac{l_{d}}{r}}}\int_0^\infty\gamma^{\frac{l_{d}}{r}-1}\ln({1+\gamma})
\nonumber
\\
    &\times G_{0,1}^{1,0}\left[\Lambda_{2,d}\left(\frac{\gamma}{\mu_{r,d}}\right)^{\frac{1}{r}}\biggl | 
    \begin{array}{c}
     - \\
     0 \\ 
    \end{array}
    \right]d\gamma.
\end{align}
Now, with the help of \cite[Eq.~(8.4.6.5)]{art3} to transform $\ln({1+\gamma})$ into Meijer's G function and then applying the formula of \cite[Eq.~(2.24.1.1)]{art3}, ACC is obtained as follows:
\begin{align}
    \label{ergf}
    \mathrm{ACC}&=\frac{\Lambda_{1,d}\mu_{r,d}^{-\frac{l_{d}}{r}}}{2\ln{(2)}r^{\frac{1}{2}}(2\pi)^{\frac{r-1}{2}}}G_{2,r+2}^{r+2,1}\left[\frac{\Lambda_{2,d}^{r}}{\mu_{r,d}r^{r}}\biggl | 
    \begin{array}{c}
     \frac{-l_{d}}{r},1-\frac{l_{d}}{r}\\
     0,\frac{r-1}{r},\frac{-l_{d}}{r} \\ 
    \end{array}
    \right].
\end{align}
\textit{Asymptotic Analysis:} By employing the same identity as that used for OP and ABER, the asymptotic expression of ACC is derived as
\begin{align}
     \label{ergs}
\mathrm{ACC}^{\infty}\approx&\sum_{h=1}^{r+2}\frac{\Lambda_{1,d}\mu_{r,d}^{t_{1,h}-\frac{l_{d}}{r}-1}r^{rt_{1,h}-r-\frac{1}{2}}}{2\ln{(2)}\Lambda_{2,r}^{rt_{1,h}-r}(2\pi)^{\frac{r-1}{2}}}
\nonumber
\\
\times & \frac{\prod_{g=1;g\neq h}^{r+2}\Gamma(t_{1,h}-t_{1,g})\Gamma(2+\frac{l_{d}}{r}-t_{1,h})}{\Gamma(t_{1,h}-\frac{l_{d}}{r})}.
\end{align}

{\subsection{Average Secrecy Capacity}
ASC is defined as the average value of instantaneous secrecy capacity. This performance metric is used to evaluate the secrecy performance of a secure wireless communication. The mathematical representation of ASC can be expressed as \cite[Eq.~(14)]{f4}
\begin{align}
\label{asc}
    ASC=\int_{0}^{\infty}\frac{ F_{\gamma_{e}}(\gamma)}{1+\gamma}[1-F_{\gamma_{d}}(\gamma)]d\gamma.
\end{align}
Substituting \eqref{e2} into \eqref{asc}, we obtain
\begin{align}
\label{asc2}
    ASC=&\int_{0}^{\infty} \frac{\Lambda_{3,e}\gamma^{\frac{l_{e}}{r}}}{\sqrt{r}\mu_{r,e}^{\frac{l_{e}}{r}}}\frac{1}{1+\gamma}
G_{1,r+1}^{r,1}\left[\Lambda_{4,e}\left(\frac{\gamma}{\mu_{r,e}}\right)\biggl | 
    \begin{array}{c}
     1-\frac{l_{e}}{r} \\
     0,{\frac{r-1}{r}},\frac{-l_{e}}{r} \\ 
    \end{array}
    \right]
\nonumber
\\    
\times&\Biggl(1-\frac{\Lambda_{3,d}\gamma^{\frac{l_{d}}{r}}}{\sqrt{r}\mu_{r,d}^{\frac{l_{d}}{r}}}G_{1,r+1}^{r,1}\left[\Lambda_{4,d}\left(\frac{\gamma}{\mu_{r,d}}\right)\biggl | 
    \begin{array}{c}
     1-\frac{l_{d}}{r} \\
     0,{\frac{r-1}{r}},\frac{-l_{d}}{r} \\ 
    \end{array}
    \right]\Biggl)d\gamma.
\end{align}
We employ the identity \cite[Eq. (8.4.2.5)]{art3} to convert $\frac{1}{1+\gamma}$ into Meijer's G function and then utilize \cite[Eq. (2.24.1.1)]{art3} and \cite[Eq. (07.34.21.0081.01)]{f5}, to derive ASC as 
\begin{align}
ASC=&\frac{\Lambda_{3,e}}{\sqrt{r}\mu_{r,e}^{\frac{l_{e}}{r}}}\Biggl(G_{2,r+2}^{r+1,2}\left[\frac{\Lambda_{4,e}}{\mu_{r,e}}\biggl | 
    \begin{array}{c}
     1-\frac{l_{e}}{r},-\frac{l_{e}}{r} \\
     0,{\frac{r-1}{r}},\frac{-l_{e}}{r} \\ 
    \end{array}
    \right]-\frac{\Lambda_{3,d}}{\sqrt{r}\mu_{r,d}^{\frac{l_{d}}{r}}}
     \nonumber   
    \\  \times&G_{1,1:1,r+1:1,r+1}^{1,1:r,1:r,1}\left[
    \begin{array}{c}
     v\\
     v\\
    \end{array}
    \biggl | 
    \begin{array}{c}
     1-\frac{l_{e}}{r} \\
     t_{3} \\ 
    \end{array}
    \biggl |
    \begin{array}{c}
     1-\frac{l_{d}}{r} \\
     t_{4} \\ 
    \end{array}
    \biggl |
    \frac{\Lambda_{4,e}}{\mu_{r,e}},\frac{\Lambda_{4,d}}{\mu_{r,d}}
    \right]\Biggl),
\end{align}
where $v=-\frac{l_{d}}{r}-\frac{l_{e}}{r}$,
$t_{3}=(0,\frac{r-1}{r},\frac{-l_{e}}{r})$,
$t_{4}=(0,\frac{r-1}{r},\frac{-l_{d}}{r})$.
}
\subsection{Secrecy Outage Probability}
SOP is used to assess the system security in the presence of potential eavesdroppers by estimating the probability of the instantaneous secrecy capacity ($C_{si}$) falling below a predetermined secrecy rate ($\tau_{s}$). The SOP is mathematically expressed as \cite[Eq.~(23)]{art20}
\begin{align}
\label{sop_def}
    \mathrm{SOP} \ge \mathrm{SOP}_{L}=\int_{0}^{\infty}F_{\gamma_{d}}(\Psi\gamma)f_{\gamma_{e}}(\gamma),
\end{align}
where $\Psi=2^{\tau_{s}}$. By replacing \eqref{e1} and \eqref{e2} with \eqref{sop_def}, SOP can be expressed as
\begin{align}
    \mathrm{SOP}_{L}&= \frac{\Lambda_{1,e}\Lambda_{3,d}\Psi^{\frac{l_{d}}{r}}}{r^{\frac{3}{2}}\mu_{r,e}^{\frac{l_{e}}{r}}\mu_{r,d}^{\frac{l_{d}}{r}}}\int_{0}^{\infty}\gamma^{\frac{l_{d}+l_{e}}{r}-1}
G_{0,1}^{1,0}\left[\Lambda_{2,e}\left(\frac{\gamma}{\mu_{r,e}}\right)^{\frac{1}{r}}\biggl | 
    \begin{array}{c}
     - \\
     0 \\ 
    \end{array}
    \right]
    \nonumber
    \\
&\times G_{1,r+1}^{r,1}\left[\Lambda_{4,d}\left(\frac{\Psi\gamma}{\mu_{r,d}}\right)\biggl | 
    \begin{array}{c}
     1-\frac{l_{d}}{r} \\
     0,{\frac{r-1}{r}},\frac{-l_{d}}{r} \\ 
    \end{array}
    \right]d\gamma. 
\end{align}
Using the identity of \cite[Eq.~(2.24.1.1) ]{art3}, the closed-form expression of the SOP is derived as
\begin{align}
     \label{sop}\mathrm{SOP}_{L}&=\frac{\Lambda_{1,e}\Lambda_{3,d}\mu_{r,d}^{\frac{l_{e}}{r}}}{r\mu_{r,e}^{\frac{l_{e}}{r}}\Psi^{\frac{l_{e}}{r}}\Lambda_{4,d}^{\frac{l_{d}+l_{e}}{r}}(2\pi)^{\frac{r-1}{2}}}
     \nonumber
     \\
&\times G_{r+1,r+1}^{r+1,r}\left[\frac{\Lambda_{2,e}^{r}\mu_{r,d}}{r^{r}\Psi \Lambda_{4,d}\mu_{r,e}}\biggl | 
    \begin{array}{c}
     \varpi,\varpi-t_{4,r},\varpi-t_{4,r+1}\\
     0,{\frac{r-1}{r}},\frac{-l_{d}}{r} \\ 
    \end{array}
    \right],
\end{align}
where $\varpi=1-\frac{l_{d}}{r}-\frac{l_{e}}{r}$.
\\
\textit{Asymptotic Analysis:} Expanding the Meijer's G function using \cite[Eq.~(41)]{art8}, the asymptotic expression for $\mathrm{SOP}_{L}$ at a high SNR is obtained as
\begin{align}
   \label{sops}  
   \mathrm{SOP}_{L}^{\infty}&\approx \sum_{h=1}^{r}\frac{\Lambda_{1,e}\Lambda_{3,d}\Lambda_{2,e}^{rt_{5,h}-r}\mu_{r,d}^{t_{5,h}+\frac{l_{e}}{r}-1}}{r^{rt_{5,h}-r+1}(\mu_{r,e}\Psi)^{t_{5,h}+\frac{l_{e}}{r}-1}\Lambda_{4,d}^{t_{5,h}+\frac{l_{d}}{r}+\frac{l_{e}}{r}-1}}
     \nonumber
     \\
&\times\frac{\prod_{g=1;g\neq h}^{r}\Gamma(t_{5,h}-t_{5,g})\prod_{g=1}^{r+1}\Gamma(1+t_{4,g}-t_{5,h})}{(2\pi)^{\frac{r-1}{2}}\Gamma(1+t_{5,r+1}-t_{5,h})},
\end{align}
where $t_{5}=\left( \varpi,\varpi-t_{4,r},\varpi-t_{4,r+1}\right)$.
\section{NUMERICAL RESULTS}\label{results}
In this section, the effect of different parameters on the system performance is investigated. The graphical representations presented were based on the analytical expressions derived for the performance metrics, as demonstrated in \eqref{op}, \eqref{ber}, \eqref{erg2}, and \eqref{sop}. Moreover, mathematical expressions incorporating $10^6$ random samples per channel were validated using Monte Carlo (MC) simulations. To ensure versatility, we configured the system parameters as $N_{d}=N_{e}=2$, $\alpha_{s,d}=\alpha_{r,d}=\alpha=5.52$, $\alpha_{s,e}=\alpha_{r,e}=3.43$, $\beta_{s,d}=\beta_{r,d}=\beta=2.34$, $\beta_{s,e}=\beta_{r,e}=1.43$, $\zeta_{s,d}=\zeta_{r,d}=\zeta_{s,e}=\zeta_{r,e}=1$, $\tau_{s}=0.1$ bits/sec/Hz, $r=1$, $(p,q)=(1,1)$, and $\mu_{r,e}=30$ dB, as utilized in \cite{art2}. Note that strong, moderate, and weak turbulence conditions were set as $(\alpha, \beta)$ = ($3.43$, $1.43$), ($5.52$, $2.34$), and ($10.67$, $4.59$), respectively. In addition, an asymptotic analysis was conducted, which provides valuable insights and demonstrates a strong agreement with the analytical results, particularly in the case of high SNR scenarios.
\begin{figure}[t]
\vspace{0mm}
\centerline{\includegraphics[width=0.4\textwidth,angle =0]{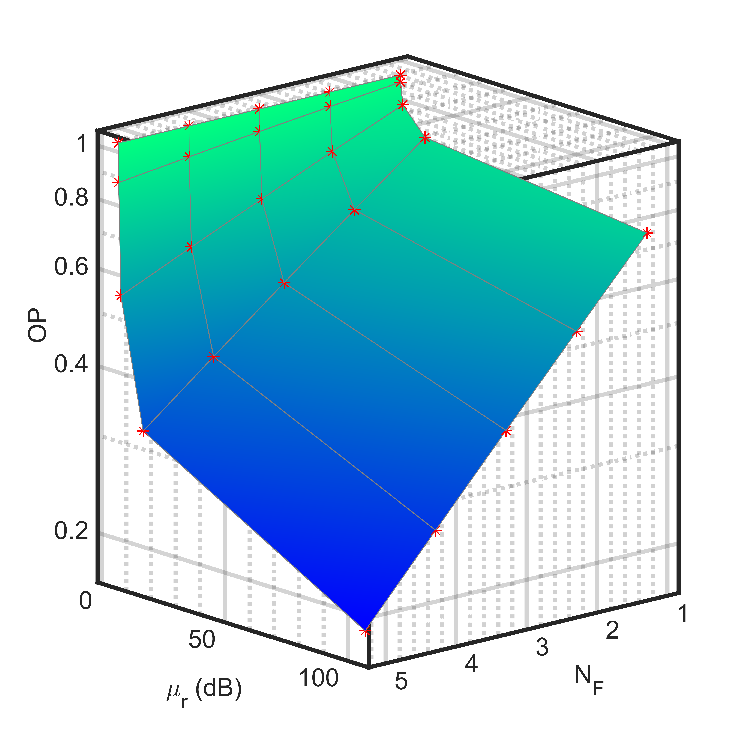}}
       \vspace{0mm}
   \caption{ OP versus $\mu_{r,d}$ and $N_{d}$.}
   \label{f1}
\end{figure}

In Fig. \ref{f1}, the graph illustrates the variation of OP with the electrical SNR ($\mu_{r,d}$) and number of reflecting elements ($N_{d}$), with a specific focus on the evaluation of the outage performance. As evident, an increase in $N_{d}$ corresponded to an improvement in the system performance. A greater number of RIS-reflecting elements facilitated more precise and focused beam-forming, resulting in enhanced signal strength and, consequently, a higher SNR at the receiver. Thus, the use of RIS is beneficial for improving the system performance. Furthermore, the illustration explored the impact of $\mu_{r,d}$, revealing that a higher electrical SNR significantly reduced the OP by ensuring superior signal quality, accelerated data transmission, and an extended communication range.
\begin{figure}[t]
\vspace{0mm}
\centerline{\includegraphics[width=0.4\textwidth,angle =0]{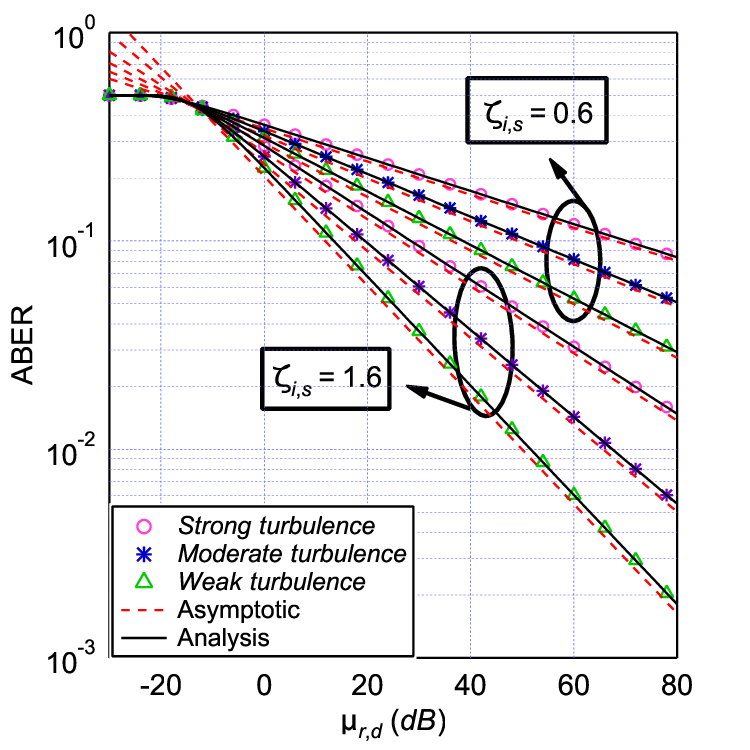}}
       \vspace{0mm}
   \caption{ABER versus $\mu_{r,d}$ for selected values of $\alpha$, $\beta$ and $\zeta_{s,d}$.}
   \label{f2}
\end{figure}

{Figures \ref{f2} - \ref{f3} provide a visual representation of the impact of various turbulence conditions on the system performance.
Increasing the turbulence parameters ($\alpha$, $\beta$) in the $\mathcal{S}-\mathcal{R_{I}}-\mathcal{U}$ link mitigated the turbulence severity, thereby improving the ABER and ASC performance. This improvement was anticipated because turbulence-induced fluctuations in the atmospheric refractive index can result in signal attenuation. As the turbulence severity increased, the optical signal losses owing to scattering and absorption were expected to increase.
\begin{figure}[t]
\vspace{0mm}
\centerline{\includegraphics[width=0.4\textwidth,angle =0]{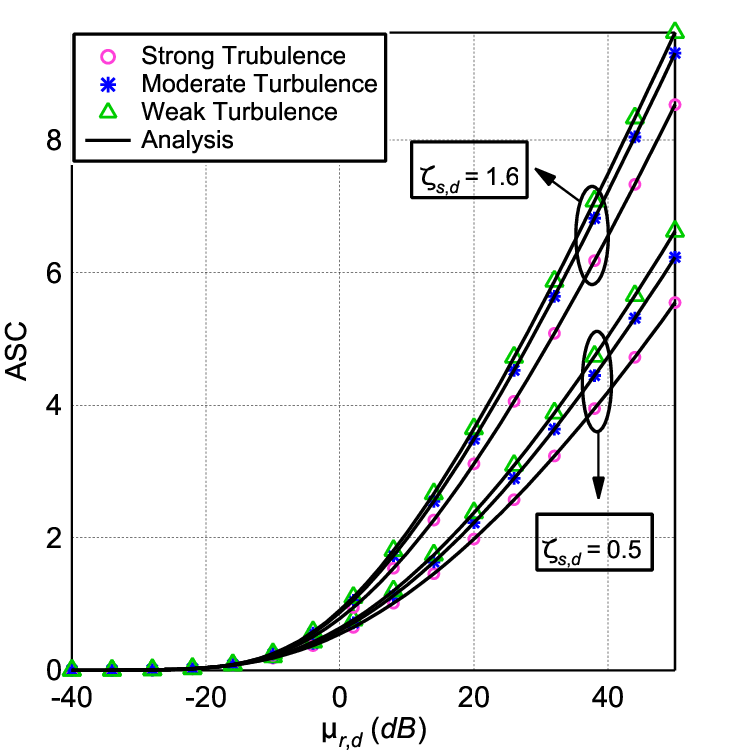}}
       \vspace{0mm}
   \caption{ASC versus $\mu_{r,d}$ for selected values of $\alpha$, $\beta$ and $\zeta_{s,d}$.}
   \label{f3}
\end{figure}
Further, these figures show the influence of the pointing error on the $\mathcal{S}-\mathcal{R_{I}}$ link. As evident, increasing $\zeta_{s,d}$ enhanced the pointing accuracy and overall system performance. This is attributed to the fact that an increase in the value of $\zeta_{s,d}$ reduced the pointing error severity resulting in a more accurate alignment of the transmitted optical beam with the receiver. This refined alignment contributed to a stronger and more stable signal, ultimately enhancing the quality of the secured signal at the receiving end.}
\begin{figure}[t]
\vspace{0mm}
\centerline{\includegraphics[width=0.4\textwidth,angle =0]{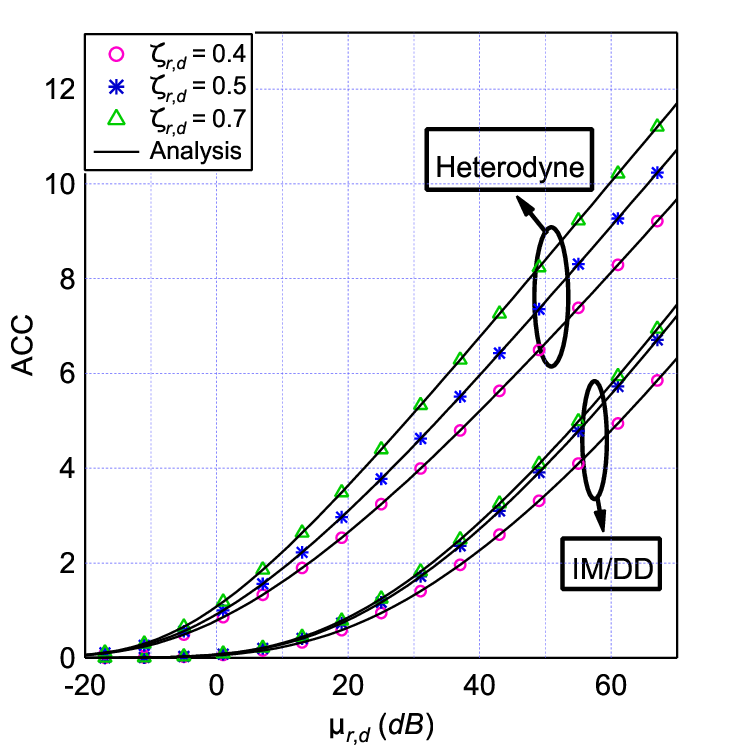}}
       \vspace{0mm}
   \caption{ ACC versus $\mu_{r,d}$ for selected values of $\zeta_{r,d}$ and $r$.}
   \label{f4}
\end{figure}
\\
To investigate the impact of the pointing error ($\zeta_{r,d}$) owing to the $\mathcal{R_{I}}-\mathcal{U}$ link, ACC was plotted against $\mu_{r,d}$ as shown in Fig. \ref{f4}. The results aligned with Fig. \ref{f2} and \ref{f3}, demonstrating that an increase in $\zeta_{r,d}$ correlated with an overall improvement in channel capacity, thereby enhancing system performance. Furthermore, the figure compares the two detection techniques, emphasizing the superior efficacy of the HD method over the IM/DD technique. This superiority was attributed to the HD technique benefiting from the sophisticated receivers capable of demodulating both amplitude and phase information.
\begin{figure}[t]
\vspace{0mm}
\centerline{\includegraphics[width=0.4\textwidth,angle =0]{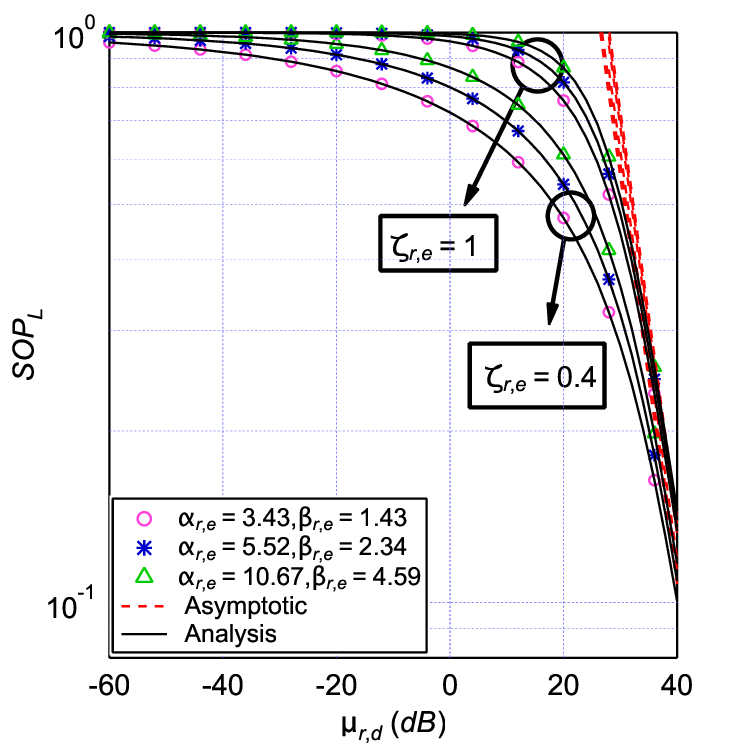}}
       \vspace{0mm}
   \caption{ $SOP_{L}$ versus $\mu_{r,d}$ for selected values of $\alpha_{r,e}$, $\beta_{r,e}$ and $\zeta_{r,e}$.}
   \label{f5}
\end{figure}
\\
Fig. \ref{f5} illustrates the impact of the turbulence parameters and pointing error on the secure outage performance of the $\mathcal{R_{I}}-\mathcal{E}$ link.
Increasing the values of ($\alpha_{r,e}$, $\beta_{r,e}$) reduced the turbulence severity within the eavesdropper channel. Consequently, this reduction increased the $SOP_{L}$ signifying the increased susceptibility of confidential data to interception or compromise. Furthermore, with the increase in $\zeta_{r,e}$, the eavesdropper could intercept information more effectively. This is because of the increased $\zeta_{r,e}$ ensuring that a greater proportion of the confidential signal reached the eavesdropper, consequently, degrading the SOP performance.
\begin{figure}[!ht]
\vspace{0mm}
\centerline{\includegraphics[width=0.4\textwidth,angle =0]{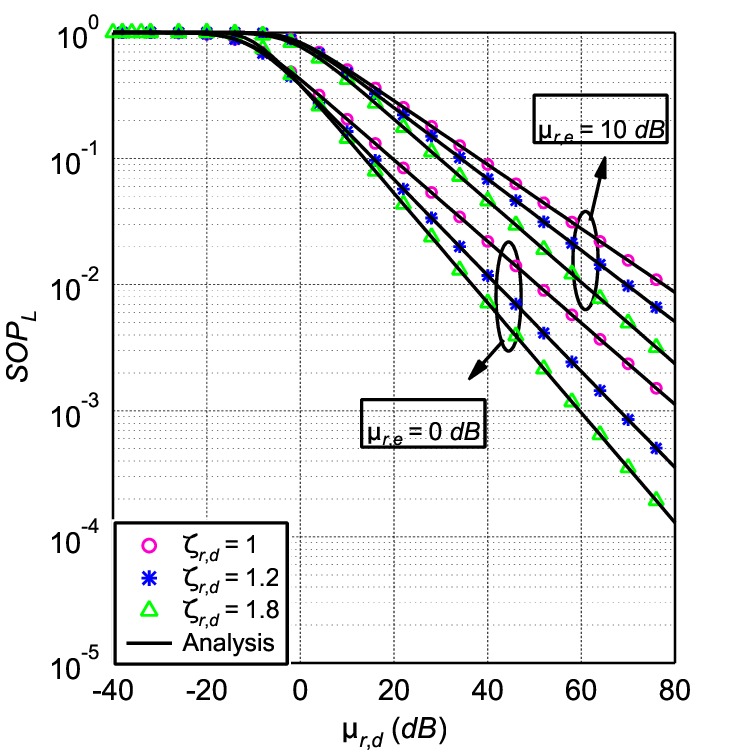}}
       \vspace{0mm}
   \caption{ $SOP_{L}$ versus $\mu_{r,d}$ for selected values of $\zeta_{r,d}$ and $\mu_{r,e}$.}
   \label{f6}
\end{figure}
\\
{To explore the impact of the pointing error ($\zeta_{r,d}$) on the secrecy performance of the proposed model, Fig. \ref{f6} illustrates the relationship between $SOP_{L}$ and $\mu_{r,d}$. The findings revealed a discernible pattern: as the pointing error parameter, $\zeta_{r,d}$, increases, the impact of pointing errors on the receiver diminishes, resulting in a reduction in $SOP_{L}$. Consequently, the optical beams were directed more precisely toward the intended recipient, thereby lowering the risk of unauthorized data access. Furthermore, a higher value of $\mu_{r,e}$ detrimentally affected the SOP performance. This is attributed to the enhanced signal strength at the eavesdropper, which facilitated easier interception of information.}
\begin{figure}[!ht]
\vspace{0mm}
\centerline{\includegraphics[width=0.4\textwidth,angle =0]{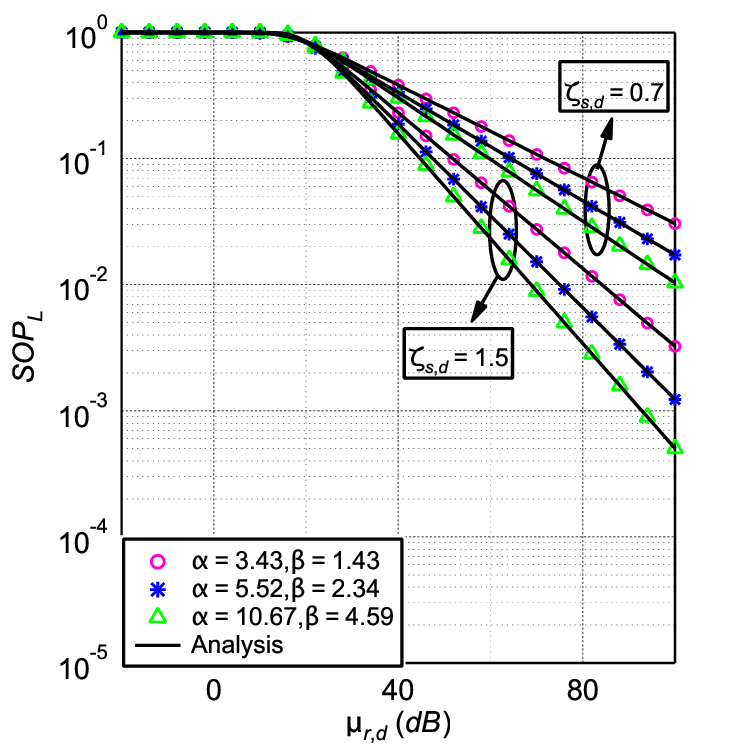}}
       \vspace{0mm}
   \caption{ $SOP_{L}$ versus $\mu_{r,d}$ for selected values of $\alpha$, $\beta$ and $\zeta_{s,d}$.}
   \label{f7}
\end{figure}
\\
{For secure FSO communication, the turbulence conditions were analyzed in Fig. \ref{f7}. As evident, the system was more secure under weak turbulence conditions ($\alpha=10.67$, $\beta=4.59$) than under moderate ($\alpha=5.52$, $\beta=2.34$) and strong turbulence conditions ($\alpha=3.43$, $\beta=1.43$). This is attributed to the higher turbulence parameters indicating less severe turbulence, resulting in minimal optical signal loss and consequently, enhanced security performance. In addition, the figure provides detailed insights into the impact of the pointing error parameter, $\zeta_{s,d}$. This observation is consistent with the results shown in Fig. \ref{f6}, where it was observed that an increase in the value of $\zeta_{s,d}$ corresponded to more secure data transfer.}

\section{CONCLUSION}
\label{co}
This study investigated the performance of an RIS-assisted FSO-based wireless network. Analytical expressions for the performance metrics were developed and validated through Monte-Carlo simulations and high SNR asymptotic analysis, which provided insights into the influence of individual parameters. The numerical findings revealed that increasing the number of reflecting elements, mitigating the turbulence severity, and minimizing the pointing errors significantly enhanced the system performance, consequently reducing the risk of wiretapping. Moreover, the results highlighted the superiority of the HD over the IM/DD technique in optical signal detection.
\label{conclusion}

\section*{APPENDIX}
\underline{\textit{FSO Turbulence Model:}}
The PDF of the fading model $\mathcal{Z}_{j}$ can be expressed as \cite[Eq.~(2)]{art2}
\begin{align}
\label{e3}
    f_{\mathcal{Z}_{j}}(x)=\frac{\Lambda_{5,j}}{x}G_{1,2}^{2,1}\left[\frac{\alpha_{j}\beta_{j}x}{\lambda_{j}-1}\biggl | 
    \begin{array}{c}
     1-\lambda_{j} \\
     \alpha_{j},\beta_{j} \\ 
    \end{array}
    \right],
\end{align}
where $\Lambda_{5,j}=\frac{1}{\Gamma(\alpha_{j})\Gamma(\beta_{j})\Gamma(\lambda_{j})}$ and $\alpha_{j}$, $\beta_{j}$, and $\lambda_{j}=\alpha_{j}-2$ denote the atmospheric turbulence parameters.

\underline{\textit{Pointing Error Model:}}
Considering the presence of a pointing error that leads to impairments, the PDF of irradiance, denoted by $I_{j}$, can be expressed as \cite[Eq.~(15)]{art1}
\begin{align} \vspace{-2mm}
    \label{e4}
    f_{I_j}(I)=\frac{\zeta_{j}^{2}}{A_{j}^{\zeta_{j}^{2}}}I^{\zeta_{j}^{2}-1},
    \vspace{-5mm}
\end{align}
where $\zeta_{j}$ denotes the pointing error at the receiver and $A_{j}$ corresponds to the pointing loss as defined in \cite{art8}.

\underline{\textit{Composite FSO Turbulence-Pointing Error Model:}}
A comprehensive statistical model based on an FSO channel was established by incorporating the 
effects of pointing errors and atmospheric turbulence. Assuming a combination of the two distributions, the PDF of $I$ is derived utilizing \cite[Eq.~(15)]{art1} as
\vspace{-1mm}
\begin{align}
\label{e5}
    f_{I'_{j}}(I)=\frac{\Lambda_{5,j}\zeta_{j}^{2}}{I}G_{2,3}^{3,1}\left[\frac{\alpha_{j}\beta_{j}I}{(\lambda_{j}-1)A_{j}}\biggl | 
    \begin{array}{c}
     1-\lambda_{j},1+\zeta_{j}^{2} \\
     \zeta_{j}^{2},\alpha_{j},\beta_{j} \\ 
    \end{array}
    \right].
\end{align}

\underline{\textit{End-to-End (E2E) SNR:}}
Let $M_{t,j}=\rho_{t,j}\varrho_{t,j}$ represent the product of two random variables, both of which follow an IGGG distribution, where $\rho_{t,j}$ denotes $\mathcal{S}-\mathcal{R}_{\mathcal{I}}$ and $\mathcal{S}-\mathcal{R}_{\mathcal{E}}$ hops with turbulence parameters $\alpha_{s,j}$, $\beta_{s,j}$, $\lambda_{s,j}$ and $\varrho_{t,j}$ signifies $\mathcal{R}_{\mathcal{I}}-\mathcal{U}$ and $\mathcal{R}_{\mathcal{E}}-\mathcal{E}$ hops with turbulence parameters $\alpha_{r,j}$, $\beta_{r,j}$, $\lambda_{r,j}$. The PDF of $M_{t,j}$ is expressed as
\begin{align}
\label{e6}
    f_{M_{t,j}}(I)=\frac{\Lambda_{6,j}\Lambda_{7,j}}{I}G_{6,4}^{2,6}\left[\frac{1}{\Lambda_{8,j}\Lambda_{9,j}I}\biggl | 
    \begin{array}{c}
     s_{1,j} \\
     s_{2,j} \\
    \end{array}
    \right],
\end{align}
where 
$\Lambda_{6,j}=\frac{\zeta_{s,j}^{2}}{\Gamma(\alpha_{s,j})\Gamma(\beta_{s,j})\Gamma(\lambda_{s,j})}$, $\Lambda_{7,j}=\frac{\zeta_{r,j}^{2}}{\Gamma(\alpha_{r,j})\Gamma(\beta_{r,j})\Gamma(\lambda_{r,j})}$, $\Lambda_{8,j}=\frac{\alpha_{s,j}\beta_{s,j}}{A_{s,j}(\lambda_{s,j}-1)}$, $\Lambda_{9,j}=\frac{\alpha_{r,j}\beta_{r,j}}{A_{r,j}(\lambda_{r,j}-1)}$, $s_{1,j}=\left[1-\zeta_{r,j}^{2},1-\alpha_{r,j},1-\beta_{r,j},1-\zeta_{i,s}^{2},1-\alpha_{i,s},1-\beta_{i,s}\right]$, and $s_{2,j}=\left[\lambda_{r,j},\lambda_{i,s},-\zeta_{i,s},-\zeta_{r,j}\right]$. Now, utilizing the first term of Laguerre expansion, the PDF of $\Bar{M}_{t,j}$ is obtained as
\begin{align}
\label{e7}
   f_{\Bar{M}_{t,j}}(y)=\frac{y^{l_{j}-1}}{\Gamma(l_{j})k_{j}^{l_{j}}}G_{0,1}^{1,0}\biggl[\frac{y}{k_{j}}\biggl | 
    \begin{array}{c}
     -\\
     0 \\ 
    \end{array}
    \biggl].
\end{align}
where $\Bar{M}_{t,j}=\sum_{t=0}^{N_{j}}M_{t,j}$ indicates the sum of positive random variables of $M_{t,j}$. The $k^{th}$ moment of the random variable, $M_{t,j}$ is obtained as

\begin{align}
    \mathbb{E}(M_{t,j}^{k})=\frac{\Lambda_{6,j}\Lambda_{7,j}}{(\Lambda_{8,j}\Lambda_{9,j})^{k}}\frac{\prod_{n=1}^{6}\Gamma(R_{mn}+k)\prod_{n=1}^{2}\Gamma(1\!\!-\!\!P_{mn}\!\!-\!\!k)}{\prod_{n=3}^{4}\Gamma(P_{mn}+k)},
    \vspace{-5mm}
\end{align}
where $P_{mn}=\begin{bmatrix}
    1-\lambda_{r,j} & 1-\lambda_{s,j} & 1+\zeta_{s,j}^{2} & 1+\zeta_{r,j}^{2}
\end{bmatrix}$ and
$R_{mn}=\begin{bmatrix}
    \zeta_{r,j}^{2} & \alpha_{r,j} & 
    \beta_{r,j} & 
    \zeta_{s,j}^{2} & \alpha_{s,j} & 
    \beta_{s,j}
\end{bmatrix}$. 
Through the utilization of random variable transformation and subsequent mathematical operations, we can demonstrate the PDF of $\gamma_{j}$ as outlined in Eq. ($1$). Now, utilizing \cite[Eq.~(9)]{art9}, the CDF of $\gamma_{j}$ is demonstrated 
as shown in \eqref{e2}. This concludes the proof of \textit{LEMMA 1}.






\bibliographystyle{IEEEtran}
\bibliography{IEEEabrv,main}

\begin{thebibliography}{10}
\providecommand{\url}[1]{#1}
\csname url@samestyle\endcsname
\providecommand{\newblock}{\relax}
\providecommand{\bibinfo}[2]{#2}
\providecommand{\BIBentrySTDinterwordspacing}{\spaceskip=0pt\relax}
\providecommand{\BIBentryALTinterwordstretchfactor}{4}
\providecommand{\BIBentryALTinterwordspacing}{\spaceskip=\fontdimen2\font plus
\BIBentryALTinterwordstretchfactor\fontdimen3\font minus \fontdimen4\font\relax}
\providecommand{\BIBforeignlanguage}[2]{{%
\expandafter\ifx\csname l@#1\endcsname\relax
\typeout{** WARNING: IEEEtran.bst: No hyphenation pattern has been}%
\typeout{** loaded for the language `#1'. Using the pattern for}%
\typeout{** the default language instead.}%
\else
\language=\csname l@#1\endcsname
\fi
#2}}
\providecommand{\BIBdecl}{\relax}
\BIBdecl

\bibitem{art9}
M.~Ibrahim, A.~Badrudduza, M.~S. Hossen, M.~K. Kundu, I.~S. Ansari, and I.~Ahmed, ``On effective secrecy throughput of underlay spectrum sharing $\alpha-\mu$/málaga hybrid model under interference-and-transmit power constraints,'' \emph{IEEE Photonics Journal}, vol.~15, no.~2, pp. 1--13, 2023.

\bibitem{ris1}
W.~Khalid, Z.~Kaleem, R.~Ullah, T.~Van~Chien, S.~Noh, and H.~Yu, ``Simultaneous transmitting and reflecting-reconfigurable intelligent surface in 6g: Design guidelines and future perspectives,'' \emph{IEEE Network}, 2022.

\bibitem{ris2}
W.~Khalid, H.~Yu, J.~Cho, Z.~Kaleem, and S.~Ahmad, ``Rate-energy tradeoff analysis in ris-swipt systems with hardware impairments and phase-based amplitude response,'' \emph{IEEE Access}, vol.~10, pp. 31\,821--31\,835, 2022.

\bibitem{ris3}
S.~Noh, J.~Lee, G.~Lee, K.~Seo, Y.~Sung, and H.~Yu, ``Channel estimation techniques for ris-assisted communication: Millimeter-wave and sub-thz systems,'' \emph{IEEE Vehicular Technology Magazine}, vol.~17, no.~2, pp. 64--73, 2022.

\bibitem{ris4}
T.~Van~Chien, L.~T. Tu, W.~Khalid, H.~Yu, S.~Chatzinotas, and M.~Di~Renzo, ``Ris-assisted wireless communications: Long-term versus short-term phase shift designs,'' \emph{IEEE Transactions on Communications}, 2023.

\bibitem{ris-pls1}
W.~Khalid, M.~A.~U. Rehman, T.~Van~Chien, Z.~Kaleem, H.~Lee, and H.~Yu, ``Reconfigurable intelligent surface for physical layer security in 6g-iot: Designs, issues, and advances,'' \emph{IEEE Internet of Things Journal}, 2023.

\bibitem{ris-pls2}
J.~Bae, W.~Khalid, A.~Lee, H.~Lee, S.~Noh, and H.~Yu, ``Overview of ris-enabled secure transmission in 6g wireless networks,'' \emph{Digital Communications and Networks}, 2024.

\bibitem{ris-pls3}
W.~Khalid, H.~Yu, D.-T. Do, Z.~Kaleem, and S.~Noh, ``Ris-aided physical layer security with full-duplex jamming in underlay d2d networks,'' \emph{IEEE Access}, vol.~9, pp. 99\,667--99\,679, 2021.

\bibitem{6g}
M.~Shahjalal, W.~Kim, W.~Khalid, S.~Moon, M.~Khan, S.~Liu, S.~Lim, E.~Kim, D.-W. Yun, J.~Lee \emph{et~al.}, ``Enabling technologies for ai empowered 6g massive radio access networks,'' \emph{ICT Express}, vol.~9, no.~3, pp. 341--355, 2023.

\bibitem{art12}
V.~K. Chapala and S.~M. Zafaruddin, ``Unified performance analysis of reconfigurable intelligent surface empowered free-space optical communications,'' \emph{IEEE Transactions on Communications}, vol.~70, no.~4, pp. 2575--2592, 2021.

\bibitem{art14}
L.~Han, X.~Liu, Y.~Wang, and X.~Hao, ``Analysis of ris-assisted fso systems over f turbulence channel with pointing errors and imperfect csi,'' \emph{IEEE Wireless Communications Letters}, vol.~11, no.~9, pp. 1940--1944, 2022.

\bibitem{art13}
P.~Agheli, H.~Beyranvand, and M.~J. Emadi, ``High-speed trains access connectivity through ris-assisted fso communications,'' \emph{Journal of Lightwave Technology}, vol.~40, no.~21, pp. 7084--7094, 2022.

\bibitem{f1}
Y.~Ai, A.~Mathur, G.~D. Verma, L.~Kong, and M.~Cheffena, ``Comprehensive physical layer security analysis of fso communications over m{\'a}laga channels,'' \emph{IEEE Photonics Journal}, vol.~12, no.~6, pp. 1--17, 2020.

\bibitem{f2}
Y.~Ai, A.~Mathur, L.~Kong, and M.~Cheffena, ``Secure outage analysis of fso communications over arbitrarily correlated m{\'a}laga turbulence channels,'' \emph{IEEE Transactions on Vehicular Technology}, vol.~70, no.~4, pp. 3961--3965, 2021.

\bibitem{f3}
G.~D. Verma, A.~Mathur, Y.~Ai, and M.~Cheffena, ``Secrecy performance of fso communication systems with non-zero boresight pointing errors,'' \emph{IET Communications}, vol.~15, no.~1, pp. 155--162, 2021.

\bibitem{pls_ed}
W.~Khalid, H.~Yu, R.~Ali, and R.~Ullah, ``Advanced physical-layer technologies for beyond 5g wireless communication networks,'' p. 3197, 2021.

\bibitem{art16}
L.~Yang, J.~Yang, W.~Xie, M.~O. Hasna, T.~Tsiftsis, and M.~Di~Renzo, ``Secrecy performance analysis of ris-aided wireless communication systems,'' \emph{IEEE Transactions on Vehicular Technology}, vol.~69, no.~10, pp. 12\,296--12\,300, 2020.

\bibitem{art17}
Y.~M. Shishter, R.~Young, and F.~H. Ali, ``Irradiance probability density function for turbulence induced fading in free space optics,'' \emph{JOSA A}, vol.~39, no.~7, pp. 1267--1274, 2022.

\bibitem{art1}
A.~Jurado-Navas, J.~M. Garrido-Balsells, J.~F. Paris, M.~Castillo-V{\'a}zquez, and A.~Puerta-Notario, ``Impact of pointing errors on the performance of generalized atmospheric optical channels,'' \emph{Optics Express}, vol.~20, no.~11, pp. 12\,550--12\,562, 2012.

\bibitem{10313311}
M.~M. Rahman, A.~Badrudduza, N.~A. Sarker, M.~Ibrahim, I.~S. Ansari, and H.~Yu, ``Ris-aided mixed rf-fso wireless networks: Secrecy performance analysis with simultaneous eavesdropping,'' \emph{IEEE Access}, vol.~11, pp. 126\,507--126\,523, 2023.

\bibitem{stefanovic2021performance}
C.~Stefanovic, M.~Morales-C{\'e}spedes, and A.~G. Armada, ``Performance analysis of ris-assisted fso communications over fisher--snedecor f turbulence channels,'' \emph{Applied Sciences}, vol.~11, no.~21, p. 10149, 2021.

\bibitem{art3}
A.~P. Prudnikov, A.~Brychkov, and O.~I. Marichev, \emph{Integrals and series: special functions}.\hskip 1em plus 0.5em minus 0.4em\relax CRC Press, 1986, vol.~2.

\bibitem{ar}
M.~A. Rakib, M.~Ibrahim, A.~Badrudduza, I.~S. Ansari, S.~Chakravarty, I.~Ahmed, and S.~A. Razzak, ``A ris empowered thz-uwo relay system for air-to-underwater mixed network: Performance analysis with pointing errors,'' \emph{IEEE Internet of Things Journal}, 2024.

\bibitem{art8}
I.~S. Ansari, F.~Yilmaz, and M.-S. Alouini, ``Performance analysis of free-space optical links over málaga (m) turbulence channels with pointing errors,'' \emph{IEEE Transactions on Wireless Communications}, vol.~15, no.~1, pp. 91--102, 2015.

\bibitem{art6}
S.~Li, L.~Yang, D.~B. da~Costa, M.~Di~Renzo, and M.-S. Alouini, ``On the performance of ris-assisted dual-hop mixed rf-uwoc systems,'' \emph{IEEE Transactions on Cognitive Communications and Networking}, vol.~7, no.~2, pp. 340--353, 2021.

\bibitem{f4}
A.~Badrudduza, M.~Ibrahim, S.~R. Islam, M.~S. Hossen, M.~K. Kundu, I.~S. Ansari, and H.~Yu, ``Security at the physical layer over gg fading and megg turbulence induced rf-uowc mixed system,'' \emph{IEEE Access}, vol.~9, pp. 18\,123--18\,136, 2021.

\bibitem{f5}
S.~Wolfram \emph{et~al.}, ``The mathematica{\textregistered} book, version 4,'' \emph{Cambridge university press1999}, 1999.

\bibitem{art20}
M.~R.~A. Ruku, M.~Ibrahim, A.~Badrudduza, I.~S. Ansari, W.~Khalid, and H.~Yu, ``Effects of co-channel interference on ris empowered wireless networks amid multiple eavesdropping attempts,'' \emph{ICT Express}, 2023.

\bibitem{art2}
O.~S. Badarneh, F.~El~Bouanani, F.~Almehmadi, and H.~S. Silva, ``Fso communications over doubly inverted gamma-gamma turbulence channels with nonzero-boresight pointing errors,'' \emph{IEEE Wireless Communications Letters}, 2023.

\end{thebibliography}

\end{document}